\documentclass{PoS}

\usepackage{graphicx}
\usepackage{bm}
\usepackage{color}

\def\be{\begin{equation}}
\def\ee{\end{equation}}
\def\ba{\begin{eqnarray}}
\def\ea{\end{eqnarray}}
\newcommand{\intdP}{\int\!dP}

\title{Quasiparticle anisotropic hydrodynamics for ultrarelativistic heavy-ion collisions}

\ShortTitle{Quasiparticle anisotropic hydrodynamics for ultrarelativistic heavy-ion collisions}

\author{\speaker{Mubarak Alqahtani}%
        \thanks{M.~Alqahtani would like to thank the organizers of CPOD 2017. He also would like to acknowledge  the support he has from  Imam Abdulrahman Bin Faisal University, Saudi Arabia through a PhD fellowship.  M.~Nopoush and M.~Strickland were supported by the U.S. Department of Energy, Office of Science, Office of Nuclear Physics under Award No. DE-SC0013470.  R. Ryblewski was supported by the Polish National Science Center grants No. DEC-2012/07/D/ST2/02125 and DEC-2016/23/B/ST2/00717.}\\
       Department of Physics, Kent State University, Kent, OH 44242 United States\\
      Imam Abdulrahman Bin Faisal University, Dammam 34212, Saudi Arabia\\
       E-mail: \email{malqaht3@kent.edu}}

\author{Mohammad Nopoush\\
	Department of Physics, Kent State University, Kent, OH 44242 United States}

\author{Radoslaw Ryblewski\\
	Institute of Nuclear Physics, Polish Academy of Sciences, PL-31342 Krak\'ow, Poland}

\author{Michael Strickland\\
        Department of Physics, Kent State University, Kent, OH 44242 United States\\
        E-mail: \email{mstrick6@kent.edu}}

\abstract{In this proceedings contribution, we will review the basics of quasiparticle anisotropic hydrodynamics. Then, we will present some recent phenomenological results of 3+1d quasiparticle anisotropic hydrodynamics and Pb-Pb $2.76$ TeV collisions from the ALICE collaboration. We show that 3+1d quasiparticle anisotropic hydrodynamics model is able to describe the experimental results for Pb-Pb collisions at $2.76$ TeV quite well for many observables such as the spectra, the elliptic flow, and HBT radii in many centrality classes.}

\FullConference{Critical Point and Onset of Deconfinement\\
                 7-11 August, 2017\\
                 The Wang Center, Stony Brook University, Stony Brook, NY}

\begin{document}

\section{Introduction}
The quark-gluon plasma (QGP) can be created and studied using  heavy-ion collision experiments at the Relativistic Heavy Ion Collider (RHIC) and Large Hadron Collider (LHC). Their results show that in a broad energy range the  collective behavior of the created QGP may be described within ideal or, even more precisely, with viscous relativistic hydrodynamics.  \cite{Huovinen:2001cy,Romatschke:2007mq,Ryu:2015vwa,Niemi:2011ix}. However, to properly account for the high-momentum anisotropies that exists in  the QGP, anisotropic hydrodynamics was introduced \cite{Florkowski:2010cf,Martinez:2010sc,Martinez:2012tu,Ryblewski:2012rr,Bazow:2013ifa,Nopoush:2014pfa,Nopoush:2014qba}. In the recent few years, there have been many theoretical advancements in anisotropic hydrodynamics, but with limited phenomenological comparisons. Earlier this year, we were  able to present some phenomenological comparisons using non-conformal anisotropic hydrodynamics. For recent reviews about anisotropic hydrodynamics, we refer the reader to \cite{Strickland:2014pga,Alqahtani:2017review}.

In this proceedings contribution, we will introduce anisotropic hydrodynamics and then focus on a specific approache, quasiparticle anisotropic hydrodynamics. In quasiparticle anisotropic hydrodynamics, one introduces a single finite-temperature dependent quasiparticle mass. This thermal mass can be determined using the realistic equation of state (EoS) which is provided by  lattice measurements. Then, by some modification on the conformal EoS formalism of anisotropic hydrodynamics \cite{Nopoush:2014pfa,Nopoush:2015yga} one can make a self-consistent non-conformal framework called  quasiparticle anisotropic hydrodynamics (aHydroQP). As usual, the dynamical equations are obtained by taking moments of the Boltzmann equation with the appropriate bulk variables for the quasiparticles. After obtaining the dynamical equations, we present some phenomenological comparisons between aHydroQP and ALICE collaboration data for a few observables.

\section{Anisotropic hydrodynamics}
In anisotropic hydrodynamics, one assumes the one-particle distribution function to be of the form
\be
f(x,p) = f_{\rm iso}\!\left(\frac{1}{\lambda}\sqrt{p_\mu \Xi^{\mu\nu} p_\nu}\right)+ \delta f(x,p) ,
\label{eq:genf}
\ee
where $ \lambda $ is the scale parameter which can be identified with the temperature in the isotropic equilibrium limit and $\xi^{\mu\nu}$ is the anisotropy tensor \cite{Nopoush:2014pfa}
\be
\Xi^{\mu\nu} = u^\mu u^\nu + \xi^{\mu\nu} - \Delta^{\mu\nu} \Phi \, ,
\ee
where $u^\mu$ is the fluid four-velocity, $\xi^{\mu\nu}$ is a symmetric and traceless anisotropy tensor, and $\Phi$ is the degree of freedom associated with the bulk pressure since our system in this case is non-conformal. Within leading-order anisotropic hydrodynamics considered here  one assumes $\delta f=0$, thus we ignore $\delta f$ contributions. We also assume $\xi^{\mu\nu}$ to be diagonal since the off-diagonal elements of $\xi^{\mu\nu}$ are small \cite{Strickland:2014pga}.

In the local rest frame the distribution function in Eq.~(\ref{eq:genf}) takes the following compact form 
\be
f(x,p) =  f_{\rm eq}\!\left(\frac{1}{\lambda}\sqrt{\sum_i \frac{p_i^2}{\alpha_i^2} + m^2}\right) ,
\label{eq:fform}
\ee
where $i\in \{x,y,z\}$ and $\alpha_i \equiv (1 + \xi_i + \Phi)^{-1/2} \,$. We note that by taking $\alpha_x=\alpha_y=\alpha_z=1$ and $\lambda=T$ one recovers the isotropic equilibrium distribution function which, in the classical case, is given by $f_{e\rm q}(T)= \exp(-E/T)$.
\section{Quasiparticle anisotropic hydrodynamics (aHydroQP)}
In this approach, in order to take into account the non-conformality of the QGP,  we  assume that the QGP can be described as an ensemble of massive quasiparticles with a single temperature-dependent mass $m(T)$. Since lattice QCD results \cite{Borsanyi:2010cj} provide us with the equilibrium energy density and pressure one may use the following thermodynamic identity to find $m(T)$
\be
 \epsilon _{\rm eq}+ P_{\rm eq}=T S_{\rm eq} = 4 \pi \tilde{N} T^4 \, \hat{m}_{\rm eq}^3 K_3\left( \hat{m}_{\rm eq}\right) ,
\label{eq:meq}
\ee
with $\hat{m}_{\rm eq}=m/T$. Since we will have extra terms generated by the gradients of the mass, we need to add a background field to the energy-momentum tensor to ensure thermodynamic consistency. In this case, the energy momentum tensor is defined by 
\be
T^{\mu\nu}_{\rm eq} = T^{\mu\nu}_{\rm kinetic,eq} + g^{\mu\nu} B_{\rm eq}  \, ,
\ee
where $B_{\rm eq}$ is the equilibrium background field which can be easily determined by integrating the following equation \cite{Alqahtani:2015qja}
\be
\frac{dB_{\rm eq}}{dT} 
= -4\pi \tilde{N}m^2 T K_1(\hat{m}_{\rm eq}) \frac{dm}{dT} \, ,
\ee
We refer the reader to \cite{Alqahtani:2015qja,Alqahtani:2016rth,Alqahtani:2016ayv,Alqahtani:2017jwl,Alqahtani:2017tnq} for more details about aHydroQP framework.
\begin{figure}[t!]
\includegraphics[width=0.99\linewidth]{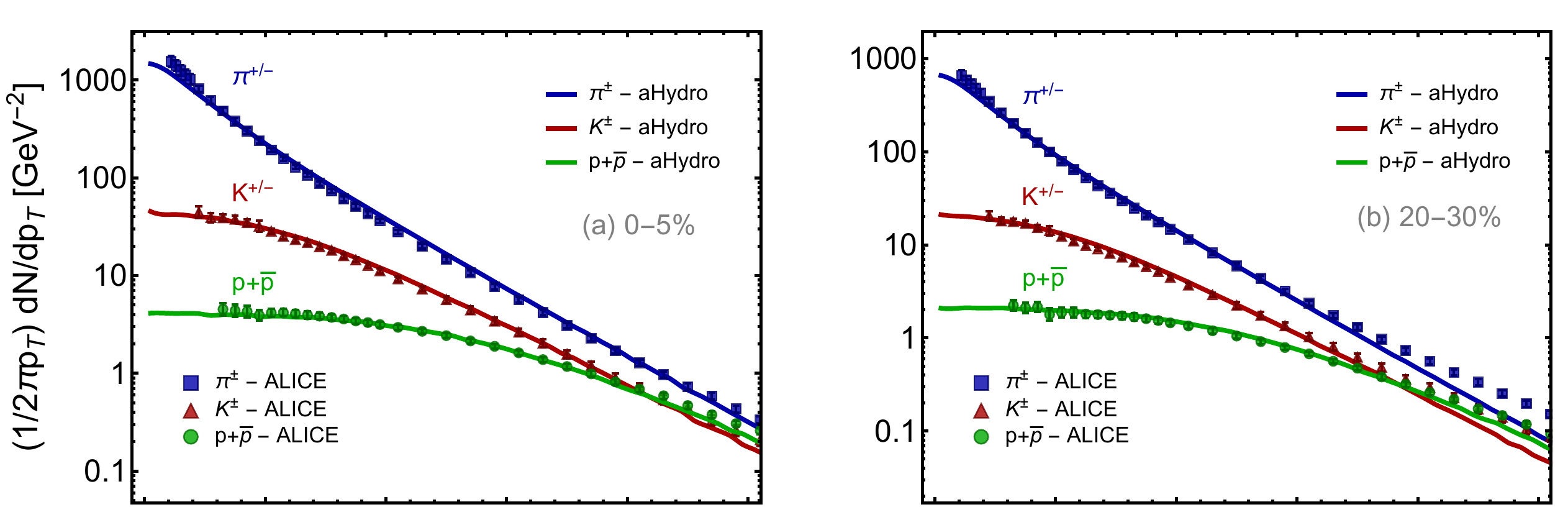}
\caption{Transverse momentum spectra of $\pi^\pm$, $K^\pm$, and $p+\bar{p}$ in 0-5\% and 20-30\% centrality classes as shown in each panel. Solid lines are aHydroQP predictions and data are taken from \cite{Abelev:2013vea}.}
\label{fig:spectra}
\end{figure}
\section{3+1d quasiparticle anisotropic hydrodynamics}
  
Next, we turn to our quasiparticle anisotropic hydrodynamics dynamical equations for  3+1d relativistic massive quasiparticle systems. To obtain the dynamical equations, one takes moments of the Boltzmann equation for particles with masses which depend on the local environment, e.g. the temperature \cite{Alqahtani:2015qja}
\be
p^\mu \partial_\mu f(x,p)+\frac{1}{2}\partial_i m^2\partial^i_{(p)} f(x,p)=-C[f(x,p)]\,,
\label{eq:boltzmanneq}
\ee
%
\begin{figure*}[t!]
\centerline{
\includegraphics[width=.49\linewidth]{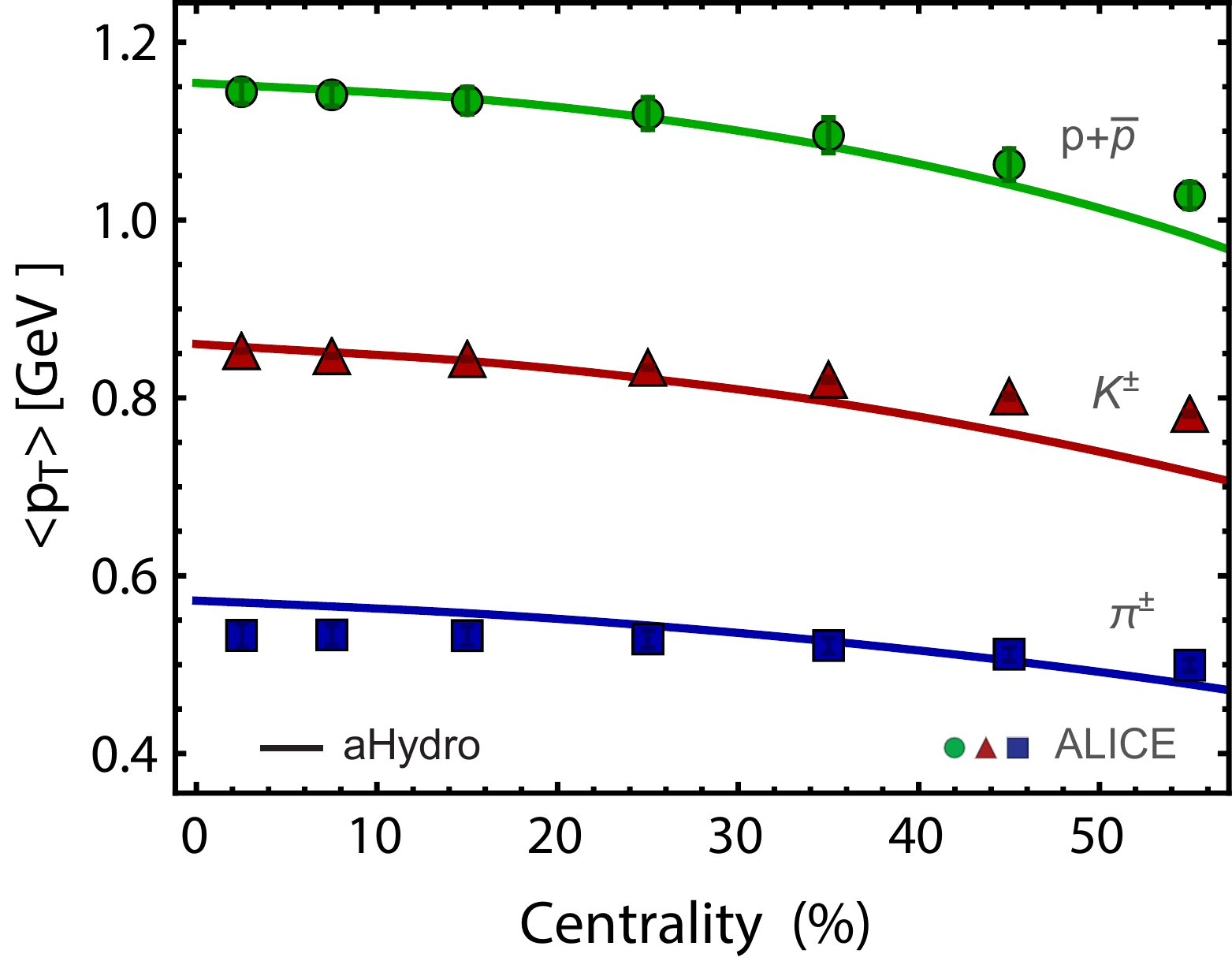}
\includegraphics[width=0.51\linewidth]{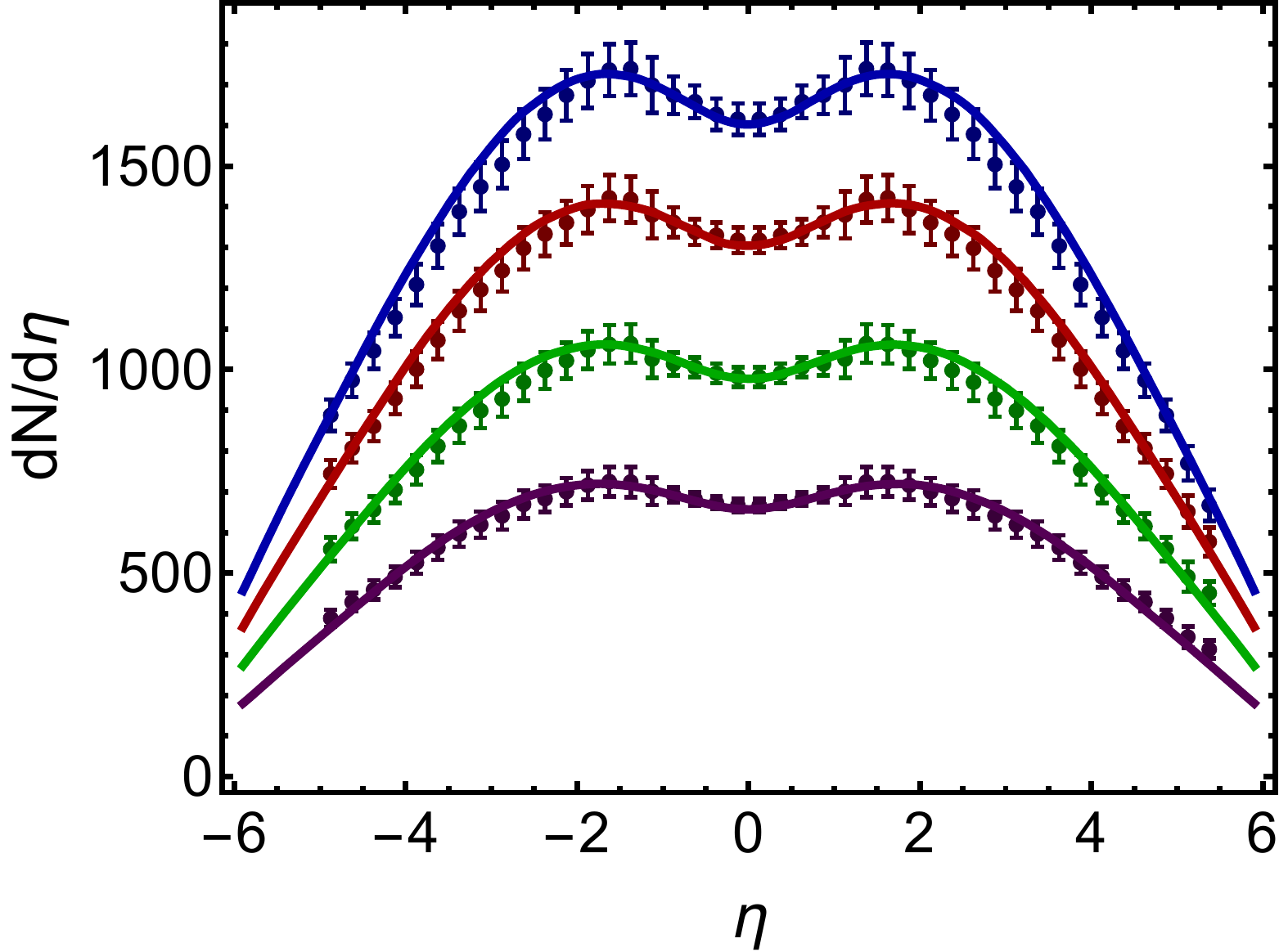}
}
\caption{In the left panel, $\langle p_T \rangle $ of pions, kaons, and protons is shown as a function of centrality. In the right panel, the charged-hadron multiplicity as a function of pseudorapidity  is shown in 0-5\%, 5-10\%, 10-20\%, and 20-30\% centrality classes.  Data are from the ALICE collaboration  Ref.~\cite{Abelev:2013vea}  and Refs.~\cite{Abbas:2013bpa,Adam:2015kda} for left panel and right panel respectively.}
\label{fig:ptavg}
\end{figure*}
where $f(x,p)$ is the anisotropic distribution function specified in Eq.~(\ref{eq:fform}) and $C[f(x,p)]$ is the collisional kernel. As we can see in the Boltzmann equation, the second term on the left-hand side corresponds to the gradients of the mass since it is a function of temperature and $T$ is a function of space-time. In practice, it suffices to take the lower-order moments, i.e., zeroth, first, and second moments to obtain the dynamical equations
\ba
\partial_\mu J^\mu&=&-\intdP \,\, C[f]\, , \label{eq:J-conservation} \\
\partial_\mu T^{\mu\nu}&=&-\intdP \, p^\nu \, C[f]\, , \label{eq:T-conservation} \\
\partial_\mu  I^{\mu\nu\lambda}- J^{(\nu} \partial^{\lambda)} m^2 &=&-\intdP \, p^\nu p^\lambda \, C[f]\, \label{eq:I-conservation},   
\ea 
where $J^\mu$, $T^{\mu\nu}$, and $ I^{\mu\nu\lambda}$ are  the particle four-current,  the energy-momentum tensor, and the rank-three tensor, respectively, defined by
\ba
J^\mu &\equiv& \intdP \, p^\mu f(x,p)\, , \label{eq:J-int} \\
T^{\mu\nu}&\equiv& \intdP \, p^\mu p^\nu f(x,p)+B({\boldsymbol\alpha},\lambda) g^{\mu\nu}, \label{eq:T-int}\\
 I^{\mu\nu\lambda} &\equiv& \intdP \, p^\mu p^\nu p^\lambda  f(x,p) \, .
\label{eq:definitions}
\ea
with $B({\boldsymbol\alpha},\lambda)$ being the non-equilibrium background field which can be found requiring thermodynamic consistency similar to the equilibrium case by this differential  equation
\be
\partial_\mu B = -\frac{1}{2} \partial_\mu m^2 \intdP  f(x,p)\,.
\label{eq:B}
\ee
Below we will list the dynamical equations in 3+1d aHydroQP. Taking the projections of the energy momentum tensor $u_\nu \partial_\mu T^{\mu \nu}$, $X_\nu \partial_\mu T^{\mu \nu}$,  $Y_\nu \partial_\mu T^{\mu \nu}$,  and $Z_\nu \partial_\mu T^{\mu \nu}$ one gets the following four equations
\ba
D_u \epsilon + \epsilon \theta_u+  P_x u_\mu D_xX^\mu+  P_y u_\mu D_yY^\mu + P_z u_\mu D_zZ^\mu &=&0\, , \\
D_x  P_x+ P_x\theta_x - \epsilon X_\mu D_uu^\mu - P_y X_\mu D_yY^\mu -  P_z X_\mu D_zZ^\mu &=& 0\,, \\
D_y  P_y+ P_y \theta_y- \epsilon Y_\mu D_uu^\mu - P_x Y_\mu D_xX^\mu -  P_z Y_\mu D_zZ^\mu  &=& 0\,,\\
D_z  P_z+ P_z \theta_z- \epsilon Z_\mu D_uu^\mu-  P_x Z_\mu D_xX^\mu -  P_y Z_\mu D_yY^\mu &=& 0\,.
\label{eq:1stmoment}
\ea
%
\begin{figure}[t!]
\includegraphics[width=0.99\linewidth]{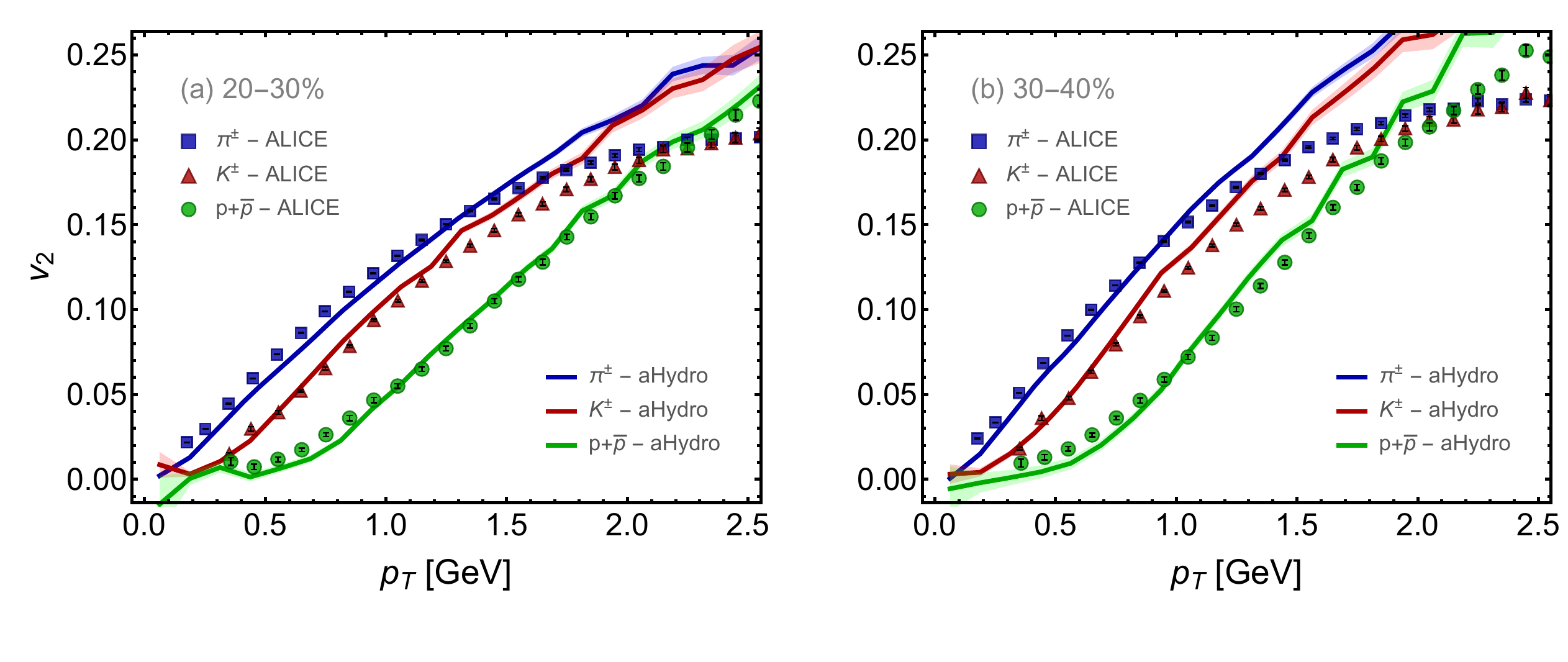}
\caption{ The identified elliptic flow coefficient as a function of  $p_T$ is shown for $\pi^\pm$, $K^\pm$, and $p+\bar{p}$ in 20-30\% and 30-40\% centrality classes. The experimental data shown  are from the ALICE collaboration ~\cite{Abelev:2014pua}.}
\label{fig:v2}
\end{figure}
where $D_u=u_\mu\partial^\mu$, $D_x=X_\mu\partial^\mu$, $D_y=Y_\mu\partial^\mu$, $D_z=Z_\mu\partial^\mu$, $\theta_u=\partial_\mu u^\mu$, $\theta_x=\partial_\mu X^\mu$, $\theta_y=\partial_\mu Y^\mu$, and $\theta_z=\partial_\mu Z^\mu$.  Another three equations are obtained from the diagonal projections of the second
moment, i,e.,  $X_\nu X_\lambda \partial_\mu I^{\mu \nu \lambda}$, $Y_\nu Y_\lambda \partial_\mu I^{\mu \nu \lambda}$, and $Z_\nu Z_\lambda \partial_\mu I^{\mu \nu \lambda}$
\ba
D_u  I_x +  I_x (\theta_u + 2 u_\mu D_x X^\mu)
&=& \frac{1}{\tau_{\rm eq}} (  I_{\rm eq} -  I_x ) \, ,  \\
D_u  I_y +  I_y (\theta_u + 2 u_\mu D_y Y^\mu)
&=& \frac{1}{\tau_{\rm eq}} (  I_{\rm eq} -  I_y ) \, , \\
D_u  I_z +  I_z (\theta_u + 2 u_\mu D_z Z^\mu)
&=& \frac{1}{\tau_{\rm eq}} (  I_{\rm eq} -  I_z ) \, .
\label{eq:2ndmoment}
\ea
Finally, one can use  the matching condition to get the last equation which relates the scale parameter $\lambda$ and the effective temperature $T$
\be
 \epsilon _{\rm kinetic}(\lambda) =  \epsilon _{\rm kinetic,eq} (T)\, .
\ee
For the definitions of $I_u$, $I_i$, $I_{\rm eq}$, one can refer to Ref.~\cite{Alqahtani:2015qja}.
\section{Phenomenological results}

We now turn to the phenomenological results. We show comparisons between 3+1d quasiparticle anisotropic hydrodynamics and the experimental data from ALICE collaboration for 2.76 TeV Pb+Pb collisions. For more comparisons, see Refs.~\cite{Alqahtani:2017jwl,Alqahtani:2017tnq}. In this model we assumed that $\eta/s={\rm const}$, we used smooth Glauber initial conditions, and we used isotropic initial conditions $\alpha_i(\tau=0)=1$. To perform the hadronic freeze-out we used a customized version of THERMINATOR 2 \cite{Chojnacki:2011hb}. 
\begin{figure}[t!]
\centerline{
\hspace{-1.5mm}
\includegraphics[width=1\linewidth]{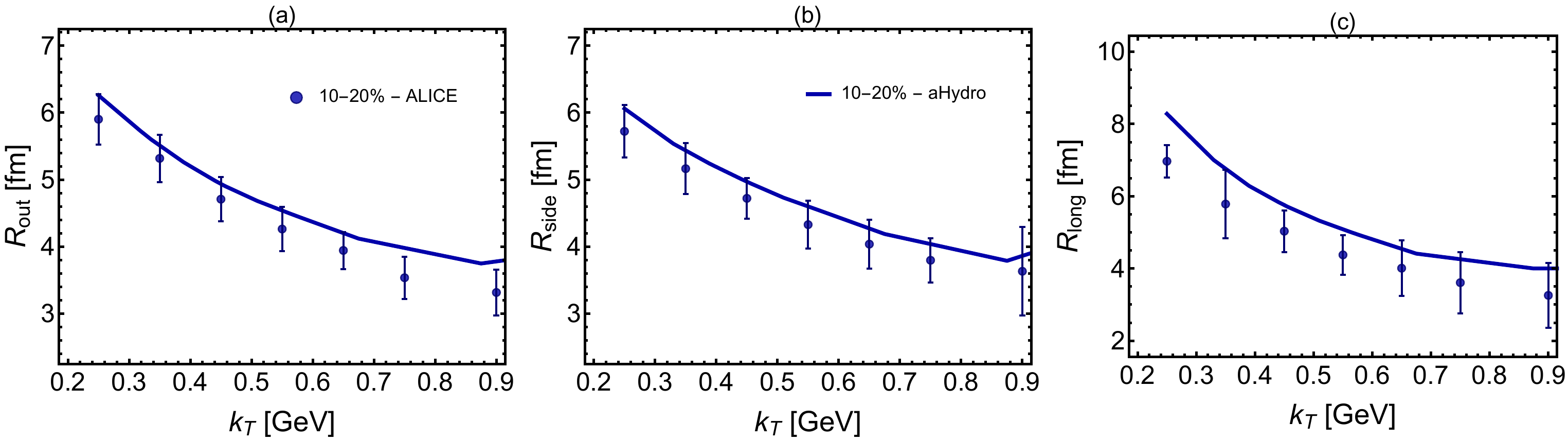}}
\centerline{
\hspace{-1.5mm}
\includegraphics[width=1\linewidth]{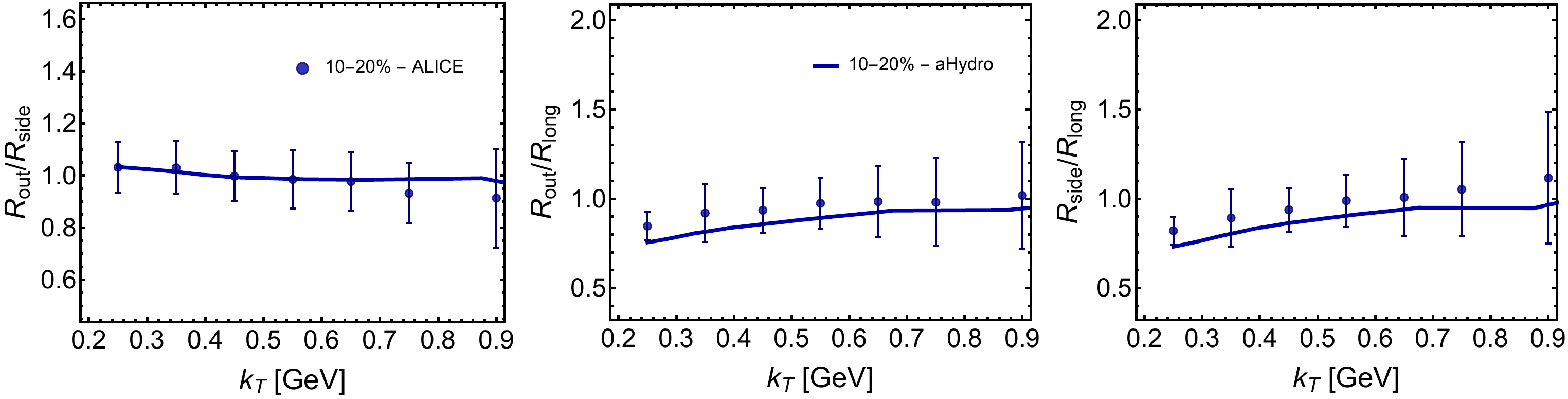}}
\caption{ The femtoscopic Hanbury-Brown-Twiss (HBT) radii as a function of the pair mean transverse  momentum for $\pi^+ \pi^+ $ is shown for 10-20\% centrality class. In the top row, the left, middle, right panels show $R_{\rm out} $, $ R_{\rm side} $, and $R_{\rm long} $ respectively. In the bottom row, the left, middle, right panels show $R_{\rm out} /R_{\rm side} $, $ R_{\rm out}/ R_{\rm long} $, and $R_{\rm side}/R_{\rm long} $ respectively. All results are for 2.76 TeV Pb+Pb collisions where  data shown are for $\pi^\pm \, \pi^\pm $  combined  from the ALICE collaboration \cite{Graczykowski:2014hoa}. }
\label{fig:HBT}
\end{figure}
Let us start by showing the spectra of pions, kaons, and protons as a function of the transverse momentum $p_T$  in Fig.~\ref{fig:spectra}. In this figure, we show  the spectra in  0-5\% and 20-30\% centrality classes. As can be seen from Fig.~\ref{fig:spectra}, our model provides a good description of  the spectra and shows the mass splitting between different hadron species. Next, in Fig.~\ref{fig:ptavg}-a, the average transverse momentum as a function of centrality is shown where aHydroQP agrees with the data quite well. In Fig.~\ref{fig:ptavg}-b, we show the multiplicity as a function of pseudorapidity in four centrality classes as shown in the figure. As we can see from this figure, aHydroQP describes the multiplicity very well compared with experimental data.

We next show the identified elliptic flow as a function of $p_T$ in two different centrality classes 20-30\% and 30-40\%  in Fig.~\ref{fig:v2}. As can be seen from this figure, our model shows a good agreement with experimental results up to $p_T \sim 1.5$ GeV. 
Finally, in the top row and bottom row of Fig.~\ref{fig:HBT}, we show comparisons of HBT radii $R_{\rm out} $, $R_{\rm side}$, and $R_{\rm long} $  and HBT radii ratios $R_{\rm out}/R_{\rm side} $, $ R_{\rm out}/R_{\rm long}  $, and $R_{\rm side}/R_{\rm long} $, respectively, as a function of the average transverse momentum. For results of other centrality classes see Ref.~\cite{Alqahtani:2017tnq}. As can be seen from this figure, aHydroQP predictions are in a good agreement with the experimental data.

\begin{figure}[t!]
\centering
\includegraphics[width=0.6\linewidth]{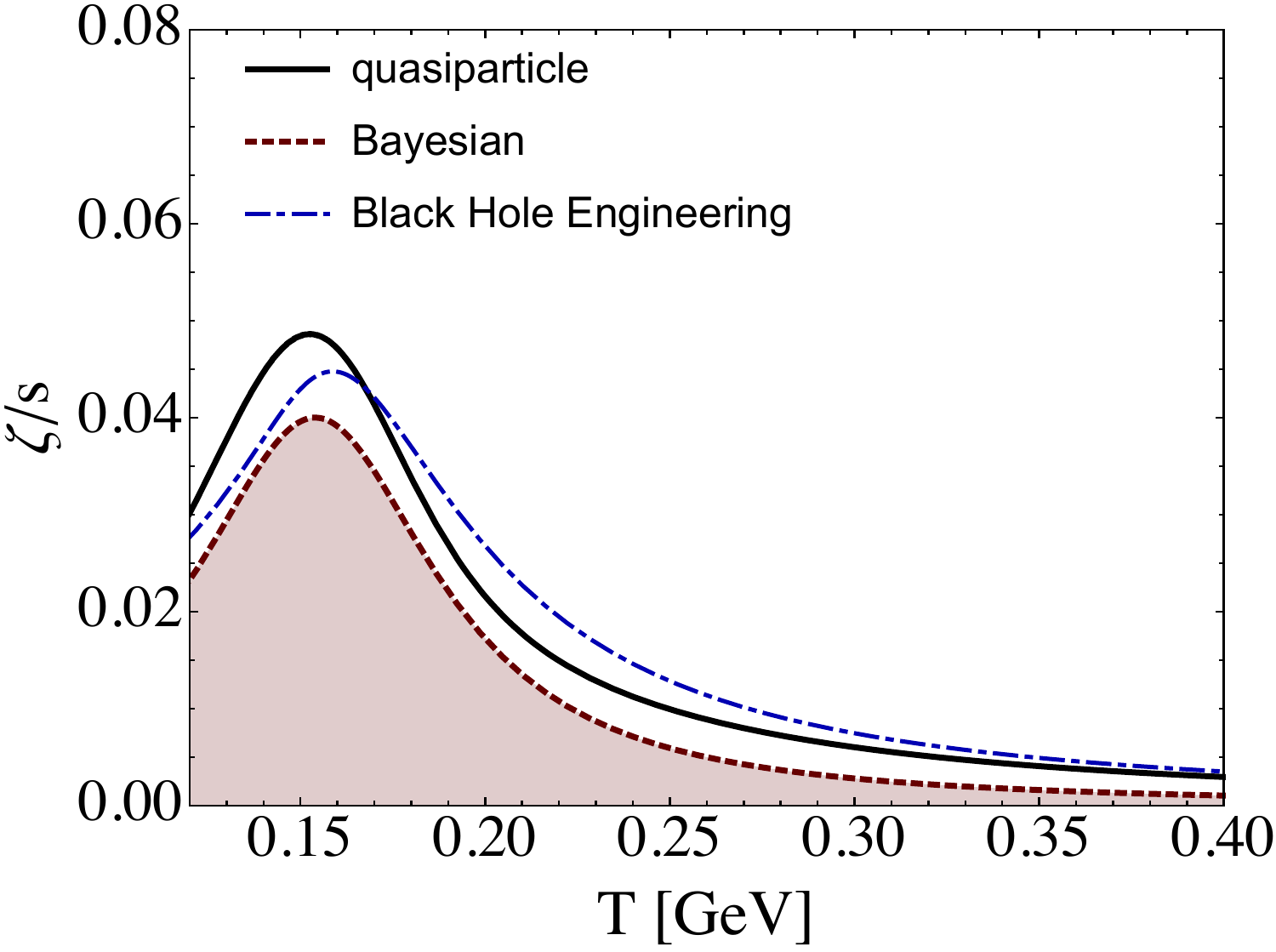}
\caption{We show the temperature dependence of the bulk viscosity scaled by the entropy density ($\zeta/S$) obtained using quasiparticle anisotropic hydrodynamics (solid-black line). For sake of comparison, we show some other model predictions: Bayesian analysis \cite{Bass:2017zyn}, and black hole engineering \cite{Rougemont:2017tlu}.}
\label{fig:zeta}
\end{figure}

Finally, we would like to list the extracted fitting parameters that we used in these comparisons which are $T_0 (\tau_0=0.25$\,fm/c$)= 600$ MeV, $\eta/s = 0.159$, and \mbox{$T_{\rm FO} = 130$ MeV}. In Fig.~\ref{fig:zeta}, we show the bulk viscosity predicted by our model aHydroQP compared with other two models: Bayesian analysis and black hole engineering where they all have good agreement ``magnitude wise''. We note here that we used $\eta/s = 0.159$ in this plot.

\section{Conclusions and outlook}
In this proceedings contribution, we showed how one can derive the dynamical equations of 3+1d quasiparticle anisotropic hydrodynamics. And how one can enforce thermodynamic consistency using the quasiparticle description in kinetic theory approaches (anisotropic hydrodynamics was an example here). We next showed some experimental comparisons with ALICE collaboration data and listed our fitting parameters extracted from fits to ALICE experimental  data. Then we showed some observables like the spectra, the mean transverse momentum as a function of centrality, the elliptic flow as a function of the mean transverse momentum, and HBT radii. In most observables, our model (aHydroQP) was able to describe the data quite well with only a few fit parameters.

Looking to the future, we are working on a few improvements to our model. For example, we are working on including temperature-dependent shear viscosity to entropy density ratio, i.e., $\eta/s(T)$. We are also planning to compare with RHIC highest energies in the near future. Moreover, we are setting up the pipeline for using fluctuating initial conditions in our code.

\bibliographystyle{JHEP}
\bibliography{MubarakCPOD}


\end{document}